\documentclass[superscriptaddress,prl,twocolumn,nofootinbib]{revtex4}
\usepackage{amsmath,amssymb}
\usepackage{graphicx,epsfig}
\usepackage{slashed}
\usepackage{float}
\newcommand{\centeron}[2]{{\setbox0=\hbox{#1}\setbox1=\hbox{#2}\ifdim
\wd1>\wd0\kern.5\wd1\kern-.5\wd0\fi \copy0
\kern-.5\wd0\kern-.5\wd1\copy1\ifdim\wd0>\wd1
                                   \kern.5\wd0\kern-.5\wd1\fi}}
\newcommand{\ltap}{\>\centeron{\raise.35ex\hbox{$<$}}
                           {\lower.65ex\hbox{$\sim$}}\>}
\newcommand{\gtap}{\>\centeron{\raise.35ex\hbox{$>$}}
                           {\lower.65ex\hbox{$\sim$}}\>}
\newcommand{\gsim}{\mathrel{\gtap}}
\newcommand{\lsim}{\mathrel{\ltap}}

\newcommand{\fref}[1]{Fig.~\ref{fig:#1}}

\newcommand{\cref}[1]{Chapter \ref{c:#1}}
\newcommand{\tref}[1]{Table~\ref{tab:#1}}

\newcommand{\BL}[1]{}
\def\beq{\begin{equation}}
\def\eeq#1{\label{#1}\end{equation}}
\def\eeqn{\end{equation}}
\def\beqa{\begin{eqnarray}}
\def\eeqa#1{\label{#1}\end{eqnarray}}
\def\eeqan{\end{eqnarray}}
\def\CR{\nonumber \\ }
\def\leqn#1{(\ref{#1})}
\newcommand{\bspace}{\!\!\!\!}
\def\gev{\mathrm{\ GeV}}
\def\tev{\mathrm{\ TeV}}
\def\pb{\mathrm{\ pb}}
\def\fb{\mathrm{\ fb}}

\def\met{\mbox{$E{\bspace}/_{T}$}}

\begin{document}

\title{SUSY-Yukawa Sum Rule at the LHC}

\author{Monika Blanke, David Curtin and Maxim Perelstein}
\affiliation{Laboratory for Elementary Particle Physics, Cornell University,
Ithaca, NY 14850}

\date{\today}

\begin{abstract}
We propose the ``supersymmetric (SUSY) Yukawa sum rule", a relationship between physical masses and mixing angles of the third-generation quarks and squarks. The sum rule follows directly from a relation between quark and squark couplings to the Higgs, enforced by SUSY. It is exactly this relation that ensures the cancellation of the one-loop quadratic divergence in the Higgs mass from the top sector. Testing the sum rule experimentally would thus provide a powerful { consistency} check on SUSY as the solution to the gauge hierarchy problem. { While such a test will most likely have to await a future next-generation lepton collider,} the LHC experiments may be able to make significant progress towards this goal. { If some of the terms entering the sum rule are measured at the LHC, the sum rule can be used (within SUSY framework) to put interesting constraints on the other terms, such as the mixing angles among third-generation squarks.}
We outline how the required mass measurements could be performed, and estimate the accuracy that can be achieved at the LHC.
\end{abstract}

\pacs{14.60.Pq, 98.80.Cq, 98.70.Vc}

\maketitle
{\em Introduction ---} Experiments at the Large Hadron Collider (LHC) have begun probing physics at the
TeV scale. The primary goal of these experiments is to understand the mechanism of electroweak symmetry breaking (EWSB). The Standard Model (SM) explanation of the EWSB, the Higgs mechanism, suffers from the hierarchy problem, whose solution very generically requires that new physics beyond the SM appear around the TeV scale. Among the candidate models of this new physics suggested by theorists, supersymmetry (SUSY) is perhaps the most appealing one. Simple SUSY models are consistent with experimental data, including gauge coupling unification, and fit naturally in string theory. Searches for SUSY will be one of the main directions pursued by the LHC experiments.

Assuming that some of the signatures predicted by SUSY models are seen at the LHC, the next major task for the experiments will be to determine the nature of the new particles involved, such as their masses and spins. In addition, there is a large number of couplings involving the new particles that one can attempt to measure. Among those, there is a small set of couplings that is, in our opinion, truly special, and deserves special attention. These are the couplings that ensure the cancellation of the quadratically divergent diagrams contributing to the Higgs mass parameter at one loop. Specific relations between these couplings and the SM gauge and Yukawa couplings are required to solve the hierarchy problem, and SUSY guarantees that these relations are satisfied. Testing these relations experimentally would clearly demonstrate the role of SUSY in restoring naturalness to the EWSB sector. The first goal of this paper is to suggest a simple sum rule, which follows unambiguously from one such coupling relation, and involves only physically measurable quantities. The second goal is to outline the set of measurements that would need to be performed to test this sum rule, and evaluate the prospects for these measurements at the LHC.
{ While a test of the sum rule will have to await a next-generation lepton collider, we find that the LHC may be able to measure several ingredients of the sum rule. Within the framework of SUSY, the sum rule can then be used to infer parameters, such as stop and sbottom mixing angles, which will be difficult or
impossible to measure directly.}

\begin{figure}[t]
\begin{center}
\includegraphics[width=8cm]{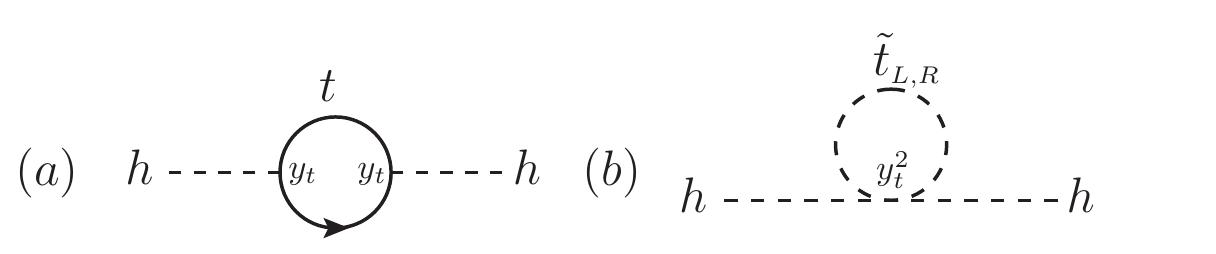}
\caption{Quadratically divergent one-loop contribution to the Higgs mass parameter in the SM (a), cancelled by scalar superpartner contributions in a SUSY model (b). }
\label{fig:loops}
\end{center}
\vspace{-8mm}
\end{figure}

{\em SUSY-Yukawa Sum Rule ---} The strongest coupling of the SM Higgs boson is the top Yukawa. At the one-loop level, this coupling introduces a quadratically divergent contribution to the Higgs mass parameter, via the diagram in Fig.~\ref{fig:loops} (a). In SUSY models, this contribution is canceled by the diagrams in \fref{loops} (b), with the {\it stops}, scalar superpartners of the top quark, running in the loop. %We would like to devise a way to test this cancelation experimentally.
The cancellation relies on the precise relation between the top Yukawa and the stop-Higgs quartic coupling, shown in the figure, which is enforced by SUSY. We would like to test this relation experimentally.
The most direct test, measuring the stop-Higgs quartic vertex, appears impossible at the LHC due to extreme smallness of all cross sections involving this vertex. However, once the Higgs gets a vacuum expectation value (vev), the quartic vertex generates a contribution to stop masses, which are in principle measurable. The challenge is to isolate this term from other contributions to the stop mass matrix.

SUSY makes two kinds of predictions: (1) it dictates a particular \emph{particle content} (i.e. superpartners to the SM fields), and (2) it imposes certain \emph{relations between couplings} of the fields, such as the relation in  Fig.~\ref{fig:loops}. We want to separate the two, fixing the particle content (which we assume could be tested by independent observations), while attempting to test the coupling relation.

Start with a SUSY-like particle content for the $3^\mathrm{rd}$ generation, i.e. a set of scalars with gauge charges
\beq
\left( \begin{array}{cc} \tilde t_L \\ \tilde b_L\end{array}\right) \sim (3,2)_{1/6}, \ \
\tilde t_R \sim (3,1)_{2/3}, \ \
\tilde b_R \sim (3,1)_{-1/3}.
\eeq{eq:qunu}
Leaving the $SU(2)_L \times U(1)_Y$ gauge symmetry unbroken and working in the $(\tilde t_L, \tilde t_R)$-basis, the only allowed mass terms are
\begin{equation}
M_{\tilde t}^2 =  \left(\begin{array}{cc}M_L^2 \\&M_t^2\end{array}\right) , \ \
M_{\tilde b}^2 =  \left(\begin{array}{cc}M_L^2 \\&M_b^2\end{array}\right)
\end{equation}
(in the MSSM these are just the soft masses). Within the chosen particle content, we can parameterize EWSB model-independently by inserting spurions $Y^{t,b}$. The $(1,1)$ entries of the top- and bottom-partner mass matrices become
\begin{equation}
\label{eqn:spurions}
(M_{\tilde t}^2)_{11} = M_L^2 + v^2 Y_{11}^t \ \ , \ \
(M_{\tilde b}^2)_{11} = M_L^2 + v^2 Y_{11}^b
\end{equation}
where $v = 246$ GeV. Let us define an observable
\beq
\Upsilon \equiv \frac{1}{v^2} \left( m_{t1}^2 c_t^2  + m_{t2}^2 s_t^2  - m_{b1}^2 c_b^2  - m_{b2}^2 s_b^2\right)\,,
\eeq{eqn:UpsilonDefinition}
where the top-partner eigenmasses $m_{t1} < m_{t2} $, the bottom-partner eigenmasses $m_{b1} < m_{b2}$, and the mixing angles $\theta_t$ and $\theta_b$ are all, in principle, measurable. (We use the notation $c_{t,b}\equiv\cos\theta_{t,b}$, $s_{t,b}\equiv\sin\theta_{t,b}$.) Writing the top-partner mass matrix in terms of these quantities:
\begin{equation}
\label{eqn:MassesAndMixings}
M_{\tilde t}^2 =\left(
\begin{array}{cc}
m_{t1}^2 c_t^2 + m_{t2}^2 s_t^2 &
 c_t s_t (m_{t1}^2 - m_{t2}^2)\\
 c_t s_t (m_{t1}^2 - m_{t2}^2) &
 m_{t1}^2 s_t^2 + m_{t2}^2 c_t^2
\end{array}\right),
\end{equation}
(similarly for $M_{\tilde b}^2$) and canceling the soft mass $M_L^2$ by evaluating $(M_{\tilde t}^2)_{11} - (M_{\tilde b}^2)_{11}$, we obtain\vspace{-2mm}
\beq
\Upsilon = Y_{11}^t - Y_{11}^b\,.
\eeq{eqn:UpsY}
In other words, $\Upsilon$ probes the spurions only. Note, however, that
Eq.~\leqn{eqn:UpsY} will receive non-trivial corrections beyond the tree level, since $\Upsilon$ is defined in terms of physical (pole) masses, while in the above derivation all masses are evaluated at the same scale.

\begin{figure}
\begin{center}
\includegraphics[width=7cm]{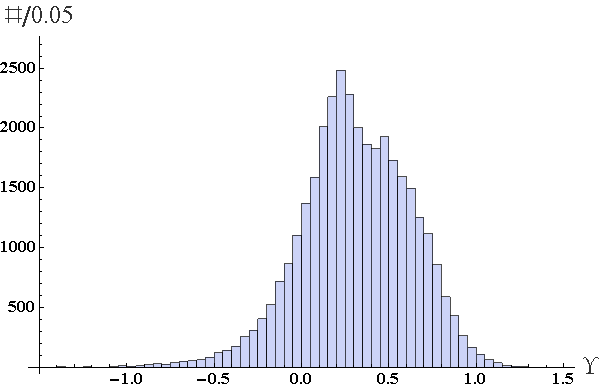}
\caption{Distribution of $\Upsilon$ for a {\texttt SuSpect} random scan of pMSSM parameter space. Scanning range was $\tan \beta \in (5,40)$; $M_A, M_1 \in (100, 500)$~GeV; $M_2, M_3, |\mu|, M_{QL}, M_{tR}, M_{bR} \in (M_1+50~{\rm GeV}, 2~{\rm TeV})$; $|A_t|, |A_b| < 1.5$ TeV; random $\mathrm{sign}(\mu)$. EWSB, neutralino LSP, and experimental constraints ($m_H, \Delta \rho$, $b\rightarrow s \gamma$, $a_\mu$, $m_{\tilde \chi^\pm_1}$ bounds) were enforced.}
\label{fig:SuSpectScan}
\end{center}
\vspace{-7mm}
\end{figure}

At tree level, SUSY makes a definite prediction for $\Upsilon$. Using the standard sfermion tree-level mass matrices (see e.g. \cite{RadiativeCorrections}) { and neglecting flavor mixing}, we obtain \vspace{-3mm}
\begin{eqnarray}
\label{eqn:SUSYprediction}
\nonumber \Upsilon^{\mathrm{tree}}_\mathrm{SUSY} &=& \frac{1}{v^2}\left( \hat m_t^2 - \hat m_b^2 + m_Z^2 \cos^2 \theta_W \cos 2 \beta \right)\\
&=& \left\{ \begin{array}{ll}0.39 & \mbox{for\ } \tan\beta = 1\\
0.28 & \mbox{for\ } \tan\beta \rightarrow \infty \end{array}\right.
\end{eqnarray}\vspace{-3mm}\\
Here the hats denote tree-level (or ``bare") masses. The numerical values assume the renormalization scale $Q=600$ GeV (so that i.e. $\hat{m}_t\approx 153$~GeV), but do not depend strongly on the precise value of $Q$.
This prediction, which we call the {\it SUSY-Yukawa sum rule}, relies on the same relation between the fermion and scalar Higgs couplings which leads to the cancelation in Fig.~\ref{fig:loops}.
%of the quadratically divergent Higgs mass correction.
Measuring $\Upsilon$ would therefore provide a powerful, if somewhat indirect method of testing whether it is SUSY that solves the hierarchy problem. (This argument is conceptually similar to the tests of the Little Higgs cancellation mechanism, proposed in \cite{LittleHiggsTest}. { Earlier examples of SUSY sum rules, devised within the mSUGRA framework, can be found in~\cite{Martin1993}.})

Radiative corrections to the SUSY prediction for $\Upsilon$ can be important, since the sum rule typically involves a rather delicate cancellation between stop and sbottom mass terms. The full analytical expressions for the radiative corrections to superpartner masses within the MSSM can be found
in~\cite{RadiativeCorrections}, and a convenient numerical implementation is provided by the {\texttt SuSpect}
package~\cite{SuSpect}. The corrections depend on a large number of MSSM parameters. To estimate their effect on $\Upsilon$, we conducted several scans of the MSSM parameter space using {\texttt SuSpect}. We did not assume a particular model of SUSY breaking, but allowed the weak-scale soft terms to vary independently. A representative result for the distribution of $\Upsilon$ is shown in Fig.~\ref{fig:SuSpectScan}. (As usual, the reader must exercise caution in interpreting this plot, since it necessarily reflects our sampling bias of parameter space.) It shows that  radiative corrections can change the value of $\Upsilon$ significantly from its tree level prediction~(\ref{eqn:SUSYprediction}). However, a measurement of $|\Upsilon| > O(1)$ would disfavor TeV-scale SUSY as the solution to the hierarchy problem. { It should be noted that in a generic theory with the particle content of Eq.~\leqn{eq:qunu}, the scalar-Higgs quartic couplings are only constrained by perturbativity, leading to the
possible range of $-16\pi^2 \lsim \Upsilon \lsim 16\pi^2$. Moreover, if some of the parameters in the sum rule are misidentified, an even broader range is possible. For example, if the mixing angle measurements were off by $\pi/2$, the right-hand side of Eq.~\leqn{eqn:UpsilonDefinition} would contain the {\it right-right} elements of the squark mass matrices, which are of course independent for stop and sbottom, so any value of $\Upsilon$ is in principle possible.
Thus, even with radiative corrections included, the SUSY-Yukawa sum rule presents a useful and
non-trivial consistency check on SUSY. }

\begin{figure*}
\begin{tabular}{ccccc}
\includegraphics[width=5.5cm]{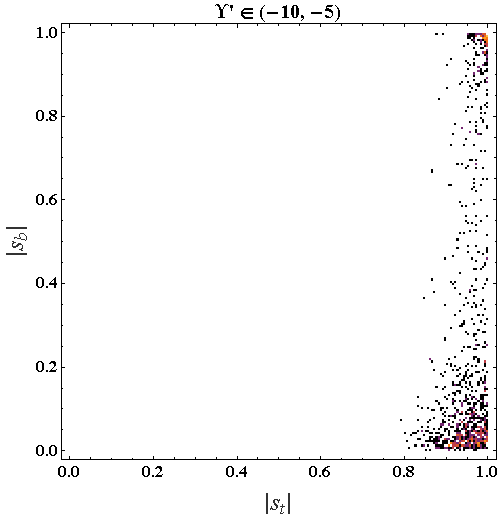}
& \phantom{bla} &
\includegraphics[width=5.5cm]{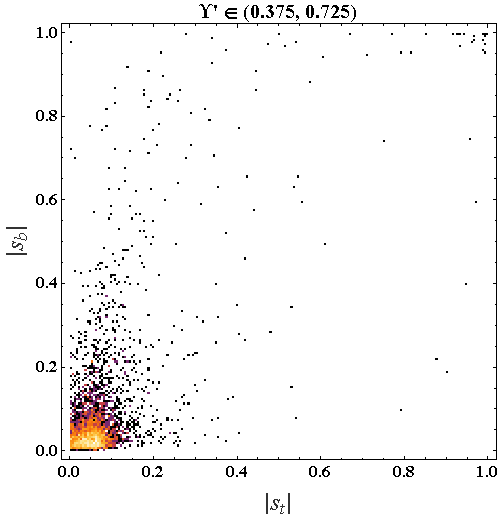}
& \phantom{bla} &
\includegraphics[width=5.5cm]{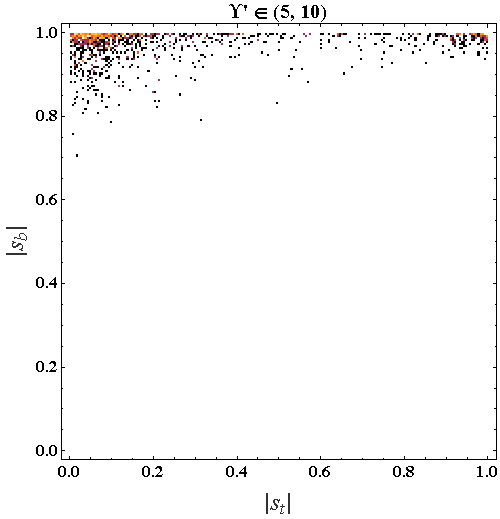}
\\
(a) &&(b)&&(c)
\end{tabular}
\caption{Scatter plot of pMSSM parameter points produced by the \texttt{SuSpect} scan from \fref{SuSpectScan}, showing the correlations between the stop and sbottom mixing angles for different ranges of $\Upsilon^\prime$. Each $0.005 \times 0.005$ bin is colored according to the number of scan points contained in it, with hot (bright) and cold (dark) colors indicating high and low scan point density, and unpopulated bins left uncolored. These correlations are a direct consequence of the SUSY-Yukawa Sum Rule, and any measurement of $\Upsilon^\prime \gsim 0$ provides valuable information about the sbottom mixing angle.}
\label{fig:angles}
\end{figure*}

{ It is also interesting to ask if the sum rule can be used as a tool for model discrimination. Recently, several SUSY ``look-alikes", {\it i.e.} models whose LHC signatures are similar to SUSY but arise from completely different underlying physics, have been studied. The most studied examples are universal
extra dimensions (UED)~\cite{UED} and little Higgs with T-parity (LHT)~\cite{LHT} models. These models contain particles with the quantum numbers of Eq.~\leqn{eq:qunu}, but instead of scalars, they are spin-1/2 fermions. (The minimal LHT model does not contain a $\tilde{b}_R$ counterpart; however, such a particle can easily be added.) This leads to a different Higgs coupling structure: for example, the 4-point coupling in Fig.~1 (b) does not exist, at renormalizable level, in these theories. As a result, UED and LHT
predictions for $\Upsilon$ are generically different from SUSY, at least at the tree level. As an example, the tree-level prediction of the minimal LHT model is
\beq
\Upsilon^\mathrm{tree}_\mathrm{LHT} = - \frac{g^\prime}{2 \sqrt{10}} \frac{m_{b_H}}{m_{A_H}} + {\cal O} \left(\frac{v^2}{f^2}\right),
\eeq{Upsilon_LHT}
where $m_{b_H}$ and $m_{A_H}$ are the masses of the heavy, T-odd partners of the left-handed $b$ quarks and the hypercharge gauge boson, respectively.
In contrast to SUSY, $\Upsilon$ is always negative at tree level in the LHT; for typical parameter values $\Upsilon\approx -0.5$. Unfortunately, radiative corrections can shift $\Upsilon$ in SUSY significantly, including changing the sign, as can be seen in Fig.~\ref{fig:SuSpectScan}. Presumably, the LHT prediction will also receive important loop corrections, although they have not yet been calculated. Depending on the resulting ranges and on the measured value of $\Upsilon$, the measurement may be interpreted as supporting one or the other model, but it seems unlikely that a sharp model-discriminating statement could be made. On the other hand, one should keep in mind that a measurement of parameters {\it not} directly entering the sum rule (such as the gluino mass) would generally shrink the range of possible $\Upsilon$ values in each model  by constraining the possible radiative corrections, improving the model-discriminating power of this observable.
}

Measuring all the ingredients of $\Upsilon$ is very difficult at a hadron collider, and the determination of the complete 3$^\mathrm{rd}$-generation sfermion spectrum and mixing angles will most likely have to be performed at a future lepton machine. However, for favorable MSSM parameters, some progress can be made at the LHC. { In particular, if some of the ingredients of the sum rule can be measured, and the sum rule is assumed to be valid, it can be used to put interesting constraints on the remaining ingredients. The easiest terms to measure at the LHC are the masses of the lightest stop and sbottom squarks. To understand the implications of such a measurement, let us rewrite $\Upsilon$ as
\begin{equation}
\Upsilon = \underbrace{\frac{1}{v^2}\left(m_{t1}^2 - m_{b1}^2\right)}_{\Upsilon^\prime}  +
 \underbrace{\frac{s_t^2}{v^2}\left(m_{t2}^2 - m_{t1}^2\right)}_{\Delta \Upsilon_t} -
 \underbrace{\frac{s_b^2}{v^2}\left(m_{b2}^2 - m_{b1}^2\right)}_{\Delta \Upsilon_b}\,.
\end{equation}
Assuming that the SUSY framework is correct, a measurement of $\Upsilon^\prime$ together with the sum rule can be used to constrain the third-generation mixing angles, even if nothing
is known about the masses of the heavier superpartners $\tilde{t}_2$ and $\tilde{b}_2$. This is illustrated by the scatter plots in Fig.~\ref{fig:angles}. If $\Upsilon^\prime$ is small, then either both $\tilde t_1$ and $\tilde b_1$ must be mostly left-handed so that $\Delta \Upsilon_{t,b}$ is small, or the two $\Delta\Upsilon$'s must precisely cancel each other. (Obviously, the second possibility is less likely, as reflected in the distribution of points in Fig.~\ref{fig:angles} (b).) A large and negative $\Upsilon^\prime$ would require a right-handed $\tilde t_1$, whereas a large and positive $\Upsilon^\prime$ requires a right-handed $\tilde b_1$. Thus, mass measurements together with the sum rule can provide non-trivial information on the mixing angles, which are difficult or impossible to measure directly at the LHC. (For some proposals for measuring the stop mixing angle, see Refs.~\cite{NojiriEdge,stop_mix}.)}

{\em Prospects at the LHC: a Case Study ---}
The MSSM parameter point we will consider is defined by the following weak-scale inputs (from here on all masses in GeV unless otherwise noted):\\
\vspace{-4mm}
\begin{center}
\begin{tabular}{|l|l|l|l|l|l|l|l|l|}
\hline
$\tan \beta$ & $M_1$ & $M_2$ & $M_3$& $\mu$ & $M_A$ &
$M_{Q3L}$ & $M_{tR}$ & $A_t$\\
\hline
 10 & 100 & 450 & 450 & 400 & 600 & 310.6 & 778.1 & 392.6\\
\hline
\end{tabular}
\end{center}
with all other $A$-terms zero and all other sfermion soft masses set at $1 \tev$. The relevant spectrum (calculated with {\texttt SuSpect}) is
\vspace{-1mm}
\begin{center}
\begin{tabular}{|l|l|l|l|l|l|l|l|}
\hline
$m_{t1}$ & $m_{t2}$ & $s_t$ & $m_{b1}$& $m_{b2}$ & $s_b$ &
$m_{\tilde g}$ & $m_{\tilde \chi^0_1}$\\
\hline
 371 & 800 & -0.095 & 341 & 1000 & -0.011& 525 & 98
 \\
\hline
\end{tabular}
\end{center}
At this benchmark point, $\Upsilon=0.423$, and $\Upsilon^\prime=0.350$. We will show below that the LHC can measure $\Upsilon^\prime$ rather accurately.

To measure the $\tilde{t}_1$ and $\tilde{b}_1$ masses, we propose to use kinematic edges, the classical $M_{T2}$ variable~ \cite{MT2original}, and recently proposed ``subsystem-$M_{T2}$" variables~\cite{MT2subsystem} to analyze the two processes \vspace{1mm}\\
\indent\begin{tabular}{lll}
(I)& $\tilde g \rightarrow  \tilde b_1 b \rightarrow b b \chi^0_1$ & via gluino pair production,\\
(II) & $\tilde t_1 \rightarrow t \chi^0_1$ & via stop pair production
\end{tabular}\vspace{1mm}\\
(where we omit antiparticle indices). For our benchmark point each of the above decays has $100\%$ branching fraction, { completely eliminating irreducible SUSY backgrounds to the measurements discussed below.}  The process (I) yields the $\tilde g, \tilde b_1$, and $\chi^0_1$ masses, and the process (II) provides $m_{t1}$. Below, we briefly outline these measurements, and estimate their accuracy; the details of this analysis will be presented in~\cite{longpaper}.

We ignore issues related to hadronization and ISR by performing the analysis at leading order in $\alpha_s$ and at parton level. %, though we briefly discuss possible extensions towards the end.
We use MadGraph/MadEvent ({ MGME}) package~\cite{MGME} to simulate gluino and stop production, and { BRIDGE}~\cite{BRIDGE} to simulate decays. We use the CTEQ6l1~\cite{cteq} parton distribution functions throughout, with the { MGME} default ($p_T$-dependent) factorization/renormalization scale choice. To roughly model detector response to jets and electrons, we introduce a Gaussian smearing of their energies according to~\cite{TDRs}\vspace{-2mm}
\begin{equation}
\frac{\Delta E_j}{E_j} = \frac{50 \%}{\sqrt {E_\gev}} \oplus 3 \% \ \ , \ \ \frac{\Delta E_e}{E_e} = \frac{10 \%}{\sqrt {E_\gev}} \oplus 0.7 \% \,.
\end{equation}\vspace{-2mm}

{\em (I) Measuring the $\tilde b_1, \tilde g, \tilde \chi^0_1$ masses --- } We study gluino pair production with subsequent decay into $4b + 2 \tilde \chi^0_1$ at the LHC with $\sqrt s = 14 \tev$ and $10 \fb^{-1}$ of integrated luminosity. The selection cuts are as follows: (a) $\met > 200 \gev$, (b) exactly 4 tagged b-jets, (c) $p_T^\mathrm{max} > 100 \gev$, (d) $p_T^\mathrm{b-jet} > 40 \gev$, (e) $|\eta| < 2.5, \Delta R > 0.4$. The gluino pair production cross section is $\sigma_{\tilde g \tilde g} \approx 11.6 \pb$. { We assumed a $b$-tag efficiency of 0.6 and b-mistag rates for $c$-, $\tau$-, and light quark/gluon jets of $0.1$, $0.1$ and $0.01$}, respectively, leaving about $1.5 \pb$ of fully $b$-tagged signal. The other kinematic cuts (a, c-e) have an efficiency of $32\%$, yielding $480 \fb$, or about $4800$ signal events at $10 \fb^{-1}$.

We computed the cross sections of the two main SM background processes, $4j+Z$ with $Z\rightarrow\nu\bar{\nu}$, and $t\bar{t}$ with one or both tops decaying leptonically. The cross sections, including efficiencies of the cuts (a-e), are $\lsim 10$~fb and
25~fb, respectively. Thus, we conclude that the SM backgrounds can be effectively eliminated by cuts, and do not take them into account further in the mass determination analysis.

\begin{figure*}
\begin{center}
\begin{tabular}{ccc}
\includegraphics[width=5cm]{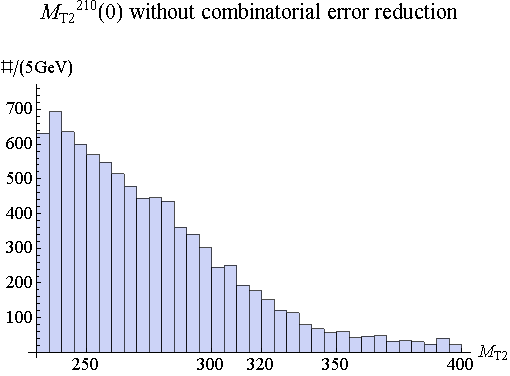}&
\includegraphics[width=5cm]{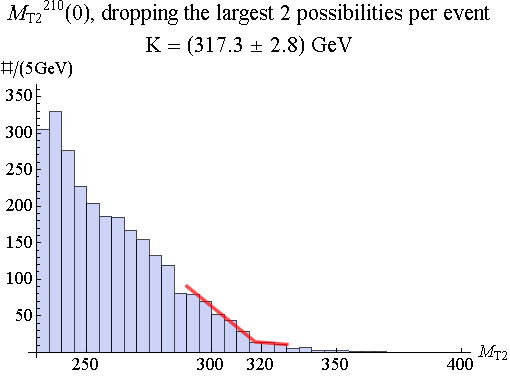}&
\includegraphics[width=5cm]{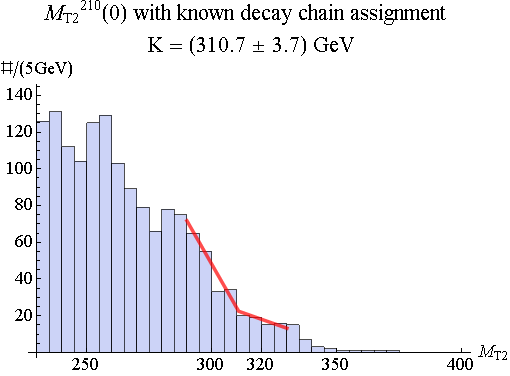}\\
(a)& (b)& (c)
\end{tabular}
\caption{$M_{T2}^{210}(0)$ distributions. The analytical prediction for the edge position is $320.9 \gev$. We emphasize that even though we show the linear kink fits only over a certain range, $K$ depends very little on the fit domain.}
\label{fig:MT2plots}
\end{center}
\vspace{-7mm}
\end{figure*}

The main background for mass determination comes from combinatorics. Consider the dijet invariant mass
$M_{bb}$. If both $b$'s come from the same decay chain, the distribution has a {\it kinematic edge} at
\beq
M_{bb}^\mathrm{max} = \sqrt{\frac{(m_{\tilde g}^2 - m_{b1}^2)(m_{b1}^2 - m_{\tilde \chi^0_1}^2)}{m_{b1}^2} }= 382.3 \gev.
\eeq{kin_edge}
For each event, there are three possible ways to assign 4 $b$'s to two decay chains, and the $M_{bb}$ distributions of the wrong combinations extend well beyond $M_{bb}^\mathrm{max}$. If all combinations are included, the edge is washed out. We find that the combinatoric background
can be reduced with simple cuts: very generally, the directions of jets from the same decay chain should be
correlated, and the pairings with the largest invariant masses are likely to be incorrect. Denoting the two $b$'s assigned to each decay chain as (1,2) and (3,4) respectively, we drop the combination with the largest $\mathrm{Max}[M_{12}, M_{34}]$ in each event, and require $\mathrm{Max}[\Delta R_{12}, \Delta R_{34}] < 2.5$. The resulting distribution shows a clear edge. We fit to it with a simple trial-PDF, the \emph{linear kink function}, which we will use throughout this analysis:
\vspace{-2mm}
\begin{center}
\includegraphics[width=2.5cm]{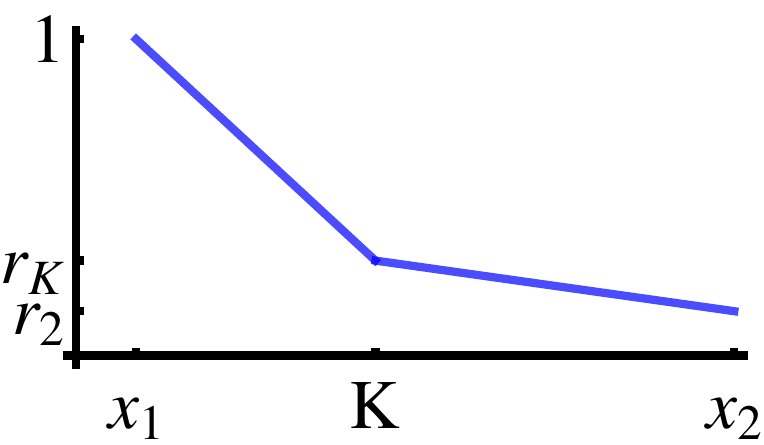}
\end{center}
\vspace{-3mm}
An unbinned maximum-likelihood fit reliably finds the edge position $K$, yielding a measurement of the kinematic edge position ${M_{bb}}_\mathrm{meas}^\mathrm{max} = (395 \pm 5) \gev$. This is quite close to the correct value, Eq.~\leqn{kin_edge}, but the use of the simple linear fit function clearly does introduce a systematic error into the edge measurement. To account for this effect, we will simply assume a systematic error of 3 times the statistical error for each edge measurement; this is  sufficient to bring across the main points of our analysis. More sophisticated methods for kinematic edge extraction exist in the literature (e.g.~\cite{NojiriEdge}), and would be used in practice.

The position of the kinematic edge provides one function of the three unknown masses; two more are required to solve for the spectrum. These can be obtained from the endpoints of distributions of events in  $M_{T2}$-subsystem variables~\cite{MT2subsystem} $M_{T2}^{220}(0)$ and $M_{T2}^{210}(0)$, predicted to be at
\beqa
{M_{T2}^{210}(0)}^\mathrm{max} &=& \frac{[(m_{b1}^2 - m_{\tilde \chi^0_1}^2)(m_{\tilde g}^2 - m_{\tilde \chi^0_1}^2)]^{1/2}}{m_{\tilde g}}= 320.9 \gev\,, \CR
{M_{T2}^{220}(0)}^\mathrm{max} &=&  m_{\tilde g} - m_{\tilde \chi^0_1}^2/m_{\tilde g} = 506.7 \gev.
\eeqa{mt2_edges}
Of the several possible $M_{T2}$ variables for this system, these two show the clearest edges, allowing precise mass determination; the complete analysis of all $M_{T2}$ variables will be presented in~\cite{longpaper}.

To calculate $M_{T2}^{210}$ for each event, we must divide the four $b$'s into an upstream and a downstream pair, giving 6 possible combinations. \fref{MT2plots} (a) shows the complete $M_{T2}^{210}(0)$ distribution; the edge is completely washed out. It turns out that of the 5 possible wrong pairings,
the two where $b$'s from the same decay chain are put into up- and down-stream pairs are the most problematic, since their $M_{T2}^{210}$ distributions extend significantly beyond the edge.
Based on this observation, we developed two techniques to reduce the combinatorial error. Firstly, for each event we can simply drop the two largest $M_{T2}$'s.
The corresponding distribution is shown in \fref{MT2plots} (b). Secondly, we can use our measurement of the kinematic edge. For each event there are three possible ways to assign the 4 $b$'s to two decay chains. For some events (about $30\%$ in our sample) we find that for two of these combinations, at least one same-chain invariant mass is larger than $M_{bb}^\mathrm{max}$, whereas for the other combination both same-chain invariant masses are smaller -- this combination must be the correct one. Using only those events and keeping only the correct decay chain assignments, we obtain the distribution of $M_{T2}^{210}(0)$ shown in \fref{MT2plots} (c). We performed linear kink fits on the distributions in \fref{MT2plots} (b) and (c), and found that they are in agreement, indicating the robustness of our approach.
Combining the two fits yields ${M_{T2}^{210}(0)}^\mathrm{max}_\mathrm{meas} =  (314.0 \pm 4.6) \gev$.
We used a similar method to extract the $M_{T2}^{220}$ edge, and obtained ${M_{T2}^{220}(0)}^\mathrm{max}_\mathrm{meas} = (492.1 \pm 4.8) \gev$.
As for the kinematic edge, the linear fit function works rather well, but it does introduce some systematic error into the edge measurements, which we again model by inflating the error bars by a factor of 3.
To summarize, the measured edges are:
\vspace{-5mm}
\begin{eqnarray}
\nonumber {M_{bb}}_\mathrm{meas}^\mathrm{max} &=& (395 \pm 15) \gev\,,\\
\nonumber {M_{T2}^{210}(0)}^\mathrm{max}_\mathrm{meas} &=& (314 \pm 14) \gev\,,\\
{M_{T2}^{220}(0)}^\mathrm{max}_\mathrm{meas} &=&  (492 \pm 14) \gev\,.
\end{eqnarray}\vspace{-5mm}\\
Each of these edges defines a subvolume of $(m_{\tilde g}, m_{\tilde \chi^0_1}, m_{b_1})$-space, which yields the mass measurements given in \tref{massmeas}.

\begin{table}
\begin{tabular}{|l|l||l|l|l|l|l|}
\hline mass & theory& median & mean & 68\% c.l.  & 95\% c.l. & process\\
\hline
$m_{b_{1}}$ & 341 & 324 & 332 & (316, 356) & (308, 432)& I\\
$m_{\tilde g}$ & 525 & 514 & 525 & (508, 552) & (500, 634) & I\\
$m_{\tilde \chi^0_1}$ & 98 &-- & -- & (45, 115) & (45, 179) & I + LEP\\
$m_{t_1}$ & 371 & 354 & 375 & (356, 414) & (352, 516) & I + II\\
\hline
\end{tabular}
\caption{Mass measurements (all in GeV), assuming Gaussian edge measurement uncertainties. We imposed the lower bound $m_{\tilde \chi^0_1} > 45 \gev$, which generically follows from the LEP invisible $Z$ decay width measurement~\cite{LEPbound}.}
\label{tab:massmeas}
\vspace{-4mm}
\end{table}

{\em (II) Measuring the $\tilde t_1$-mass --- } We simulate $p p \rightarrow \tilde t_1 \tilde t_1^* \rightarrow  t \bar t + 2 \tilde \chi^0_1$ for $100 \fb^{-1}$ integrated luminosity. The signal production cross section is $2 \pb$. The dominant irreducible background is $(Z \rightarrow \nu \nu)t \bar t$ with $\sigma_\mathrm{BG} = 135$ fb. Following~\cite{MeadeReece}, we demand two fully reconstructed hadronic tops in each event, in order to
use the classical $M_{T2}$ variable~\cite{MT2original}. Our signal cuts are (a) exactly 2 tagged b-jets and at least 4 other jets with $p_T > 30 \gev$ and $|\eta| > 2.5$ (b) lepton veto (c) $\Delta R > 0.4$ between all the b- and light jets (d) $\met > 100 \gev$ (e) $H_T > 500 \gev$ (e) $p_T^\mathrm{max} > 100 \gev$ (f) require $4j$ to reconstruct to two $W$'s with a mass window of $(60, 100) \gev$ and the two $W$'s to reconstruct with the two $b$'s to two tops with a mass window of $(140, 200) \gev$. After cuts we are left with 1481 signal and 105 background events. Plotting the classical $M_{T2}$ distribution we see a clear edge, and using the linear kink fit trial PDF with error scaling yields\vspace{-2mm}
\begin{equation}
M_{T2}(0)^\mathrm{max}_\mathrm{meas} = (340 \pm 4) \gev.
\end{equation} \vspace{-6mm}\\
Compare this to the analytical prediction \cite{MT2analytical} $
M_{T2}(0)^\mathrm{max} = 336.7 \gev$. Combined with the $m_{\tilde{\chi}^0_1}$ measurement from (I), this yields the stop mass $m_{t1}$, see \tref{massmeas}. Taking into account all correlations, we find:\vspace{-1mm}
\begin{equation}
\label{eq:Ups_meas}
\Upsilon^\prime_\mathrm{meas} = \frac{1}{v^2}\left(m_{t1}^2 - m_{b1}^2\right) = 0.525^{+ 0.20}_{- 0.15}\,,
\end{equation}
in good agreement with the theoretical value $\Upsilon^\prime=0.350$. { As explained above, a measurement of $\Upsilon^\prime$ does not by itself provide a consistency check of SUSY, or help in
discriminating it from other models. However, if the SUSY-Yukawa sum rule is assumed to be valid, this measurement can be used to place a constraint on the 3rd generation squark mixing. The measurement
in Eq.~\leqn{eq:Ups_meas} corresponds to the range of $\Upsilon^\prime$ assumed in Fig.~\ref{fig:angles} (b). Thus, even without using information from any other measurements, one could conclude that, most likely, the stop and sbottom mixing angles are rather small and the observed light stop and sbottom states are mostly left-handed (although right-handed light states, with an accidental cancellation of $\Delta\Upsilon_b$ and $\Delta\Upsilon_t$, would remain as a logical possibility at this point).}

{\em Discussion and Conclusions --- }
In this paper we proposed the SUSY Yukawa sum rule with direct connection to the cancelation of quadratic Higgs mass divergence, and introduce an observable $\Upsilon$ that can be used to test it. This constitutes a significant check on TeV-scale SUSY as the solution to the hierarchy problem. While full measurement of $\Upsilon$ will have to be left to a future lepton machine, we have demonstrated that progress could already be made at the LHC. { In particular, we showed that, for the MSSM benchmark point we chose, two masses entering the sum rule, $m_{t1}$ $m_{b1}$, can be measured. Given these measurements, one could then use the sum rule (within the SUSY framework) to put interesting constraints on other parameters, such as third-generation squark mixing angles, whose direct measurement would be difficult or impossible.}

In the course of the analysis we developed new techniques for reducing combinatorial background for $M_{T2}$-measurements, allowing for complete mass determination of $\tilde t_1, \tilde b_1, \tilde g$ and $\tilde \chi^0_1$. { At this point, we performed the analysis at the parton level, with only a crude Gaussian smearing to account for detector effects. It is important to confirm the proposed techniques with more detailed simulations including initial and final state radiation, showering and fragmentation, and better detector modeling. Results of a study including some of these effects will be presented in Ref.~\cite{longpaper}. In the future, it will also be interesting to assess the abilities of the LHC to test the sum rule (fully or partially) in the MSSM parameter regions with spectra different from our benchmark point, as well as to study in detail how the sum rule tests can be completed at a future lepton collider.}

{\em Acknowledgments ---} We are grateful to James Alexander and Konstantin Matchev for useful discussions. This work is supported by the U.S. National Science Foundation through grant PHY-0757868 and CAREER award PHY-0844667. MB thanks the Galileo Galilei Institute for Theoretical Physics for the hospitality and the INFN for partial support during the completion of this work.

\end{document}